 \documentclass[twocolumn,twocolappendix]{aastex631}

\usepackage{graphicx}
\usepackage{color}
\usepackage{rotating}
\usepackage{morefloats}
\usepackage{amssymb}

\shorttitle{ALMA number counts}
\shortauthors{Bonato et al.}

\begin{document}

\title[100 GHz ALMA number counts]{ALMACAL. XV. Band 3 ALMA Survey and Number Counts}

\author[0000-0001-9139-2342]{Matteo Bonato}
\affiliation{INAF--Istituto di Radioastronomia and Italian ALMA Regional Centre, Via Gobetti 101, Bologna, Italy, I-40129}
\email{matteo.bonato@inaf.it}

\author{Ivano Baronchelli}
\affiliation{INAF--Istituto di Radioastronomia and Italian ALMA Regional Centre, Via Gobetti 101, Bologna, Italy, I-40129}

\author{Gianfranco De Zotti}
\affiliation{INAF - Osservatorio Astronomico di Padova, Vicolo dell’Osservatorio 5, I-35122 Padova, Italy}

\author{Leonardo Trobbiani}
\affiliation{INAF--Istituto di Radioastronomia and Italian ALMA Regional Centre, Via Gobetti 101, Bologna, Italy, I-40129}
\affiliation{Dipartimento di Fisica e Astronomia, Alma Mater Studiorum Universit\`a di Bologna, via Piero Gobetti 93/2, I-40129, Bologna, Italy}

\author{Michele Delli Veneri}
\affiliation{INFN Section of Naples (INFN), Via Cinthia, I-80126, Napoli, Italy}

\author{Fabrizia Guglielmetti}
\affiliation{European Southern Observatory, Karl-Schwarzschildstrasse 2, D-85748 Garching bei München, Germany}

\author{Rosita Paladino}
\affiliation{INAF--Istituto di Radioastronomia and Italian ALMA Regional Centre, Via Gobetti 101, Bologna, Italy, I-40129}

\author{Viviana Casasola}
\affiliation{INAF--Istituto di Radioastronomia, Via Gobetti 101, Bologna, Italy, I-40129}

\author{Martin Zwann}
\affiliation{European Southern Observatory, Karl-Schwarzschildstrasse 2, D-85748 Garching bei München, Germany}

\author{Marcella Massardi}
\affiliation{INAF--Istituto di Radioastronomia and Italian ALMA Regional Centre, Via Gobetti 101, Bologna, Italy, I-40129}

\author{Elisabetta Liuzzo}
\affiliation{INAF--Istituto di Radioastronomia and Italian ALMA Regional Centre, Via Gobetti 101, Bologna, Italy, I-40129}

\author{Vincenzo Galluzzi}
\affiliation{INAF--Istituto di Radioastronomia, Via Gobetti 101, Bologna, Italy, I-40129}

\author{Erlis Ruli}
\affiliation{Dipartimento di Scienze Statistiche, Universit\`a degli Studi di Padova, Via Cesare Battisti 241/243, I-35121 Padova, Italy}

\begin{abstract}
The ALMACAL project leverages ALMA maps of calibrator-centered fields to conduct deep mm/sub-mm surveys, enabling the detection of extragalactic sources with flux densities orders of magnitude fainter than achievable with other instruments. These faint sources are critical for refining evolutionary models, as their number counts provide key constraints. In this study, we analyzed band-3 ALMACAL maps from 606 calibrator fields, employing a novel machine learning approach to mitigate the often-overlooked bias introduced by the calibrator itself. Supported by extensive simulations, we extended 100\,GHz radio AGN counts by approximately 1.5 orders of magnitude in flux density and refined constraints on dusty star-forming galaxies, reaching sensitivities as low as $\sim$180\,$\mu$Jy. We have improved the sampling, compared to previous results, in the region of the dominant population transition (between dusty star-forming galaxies and radio AGN). Our results are in good agreement with model predictions.
\end{abstract}


\section{Introduction}\label{sec:introduction}

Although substantial advancements have been achieved in the last several years, the extent of millimeter-wave (mm) surveys is still limited. Shallow all-sky surveys, reaching minimum flux densities of a few to several hundred mJy (depending on the wavelength) have been carried out by the Wilkinson Microwave Anisotropy Probe \citep[WMAP;][]{Bennett2013} and the \textit{Planck} satellite \citep{PCCS2, PCNT}. The South Pole Telescope \citep[SPT;][]{Mocanu2013, Everett2020} and the Atacama Cosmology Telescope \citep[ACT;][]{Marsden2014, Gralla2020, Vargas2024} surveys have extended the mm source counts to $\sim 10\,$mJy. Going deeper is hampered by confusion noise, even for single-dish telescopes of the 6--10\,m class like the ACT and the SPT.

The Atacama Large Millimeter/submillimeter Array (ALMA) provides the angular resolution allowing us to tear down the confusion noise, and the sensitivity to reach orders of magnitude fainter flux densities than possible with other instruments in this wavelength range, in reasonable time. 
ALMA's primary limitation in this respect is the narrow Field of View (FoV), which makes the coverage of even a small fraction of a square degree prohibitively costly in terms of observing time.

This difficulty has been overcome by the ALMA calibrators project (ALMACAL\footnote{\url{https://almacal.wordpress.com/}}), which  has calibrated and imaged the fields around  ALMA calibrators,  using data retrieved  from the ALMA archive \citep{Oteo2016}. The imaged fields have been exploited to derive deep counts of extragalactic sources in ALMA bands 6 (211--275\,GHz or 1.1--1.4\,mm), 7 (275--373\,GHz or 0.8--1.1\,mm) \citep{Oteo2016} and 8 (385--500\,GHz or 0.6--0.8  mm) \citep{Klitsch2020}. Most recently, \citet{Chen2023} exploited the most up-to-date ALMACAL database to derive number counts also in bands 3 (84--116\,GHz or 2.6--3.6\,mm), 4 (125--163\,GHz or 1.8--2.4\,mm), and 5 (158--211\,GHz or 1.4--1.9\,mm), as well as to improve those in band 6 and 7.

The ALMACAL database was also used to investigate sources of special interest \citep{Oteo2017, Klitsch2018}, to study the circumgalactic  medium of $z\le 1.4$ galaxies \citep{Klitsch2019a} and to search for extragalactic molecular absorbers \citep{Klitsch2019b}. \citet{Bonato2018} presented a catalogue of ALMA measurements of 754 calibrators observed between August 2012  and September 2017. The catalogue was extended by \citet{Bonato2019} who exploited it to validate the photometry in the \textit{Planck} multifrequency Catalogue of Non-Thermal sources \citep[PCNT;][]{PCNT} and to assess its astrometry and its completeness limits. \citet{Bonato2019} also exploited the ALMA  continuum spectra to extrapolate the observed radio source counts at 100 GHz \citep{Mocanu2013} to the effective frequencies of ALMA bands 4, 6, 7, 8, and 9 (145, 233, 285, 467, and 673 GHz, respectively).

In this paper, we revisit the ALMACAL survey in band 3. Inspecting the ALMA maps, we realized that the presence of the bright calibrator, generally a blazar, brings in a number of detections, spread over the ALMA field, that might be associated in some way to the calibrator \citep[see below and Fig.\,7 of][]{Baronchelli2024} and therefore may bias source count estimates. Most of the detected peaks cannot be serendipitous sources because they are far more numerous than expected not only from models but also from extrapolations of counts at brighter flux densities or at nearby wavelengths. Their nature is unclear. Possibilities include remnants of radio jets which in the past were oriented in different directions or residuals of the subtraction of the calibrator, or faint radio sources belonging to the over-density hosting the blazar, or something else. We note that this problem does not arise for ALMA higher frequency channels, because the radio emission sinks down rapidly with increasing frequency.

Since the peaks are distributed even at large angular distance from the calibrator, masking a small area at the field center, as usually done, is not sufficient to remove the unwanted ones. As mentioned, their nature is difficult to ascertain, also because deep multi-frequency data are missing.
To overcome this difficulty we adopted an innovative machine learning-based statistical approach which allows us not only to correct for incompleteness, contamination by spurious detections and flux boosting, but also for the bias arising from the fact that the fields are centered on bright radio sources, frequently residing in over-dense regions.

We use 100 GHz as the reference frequency of this band and 3 mm as the reference wavelength. Band 3 contains the second biggest number of ALMACAL fields, after band 6 \citep{Bonato2018}. Due to the broader FoV it covers the largest area. The full width at half maximum (FWHM) of the ALMA  primary beam at 100 GHz is $\simeq 60''$ for 12-m antennas\footnote{\url{https://almascience.nrao.edu/about-alma/alma-basics}}.



The structure of the paper is as follows. Section~\ref{sec:sample} presents a
short description of the sample. The source extraction, the corrections for incompleteness, contamination by spurious sources and flux boosting, and the removal of signals related to the calibrator are briefly described in Sect.~\ref{sec:methodology}; full details are provided in \cite{Baronchelli2024}. The results are presented in Sect.~\ref{sec:results}, while the main conclusions are summarized  in Sect.~\ref{sec:conclusions}. 

Throughout this paper we use a flat $\Lambda \rm CDM$ cosmology with $\Omega_{\rm m} = 0.31$,
$\Omega_{\Lambda} = 0.69$ and $h=H_0/100\, \rm km\,s^{-1}\,Mpc^{-1} = 0.677$
\citep{Planck2020parameters}.

\begin{figure}
\begin{center}
\includegraphics[width=\columnwidth]{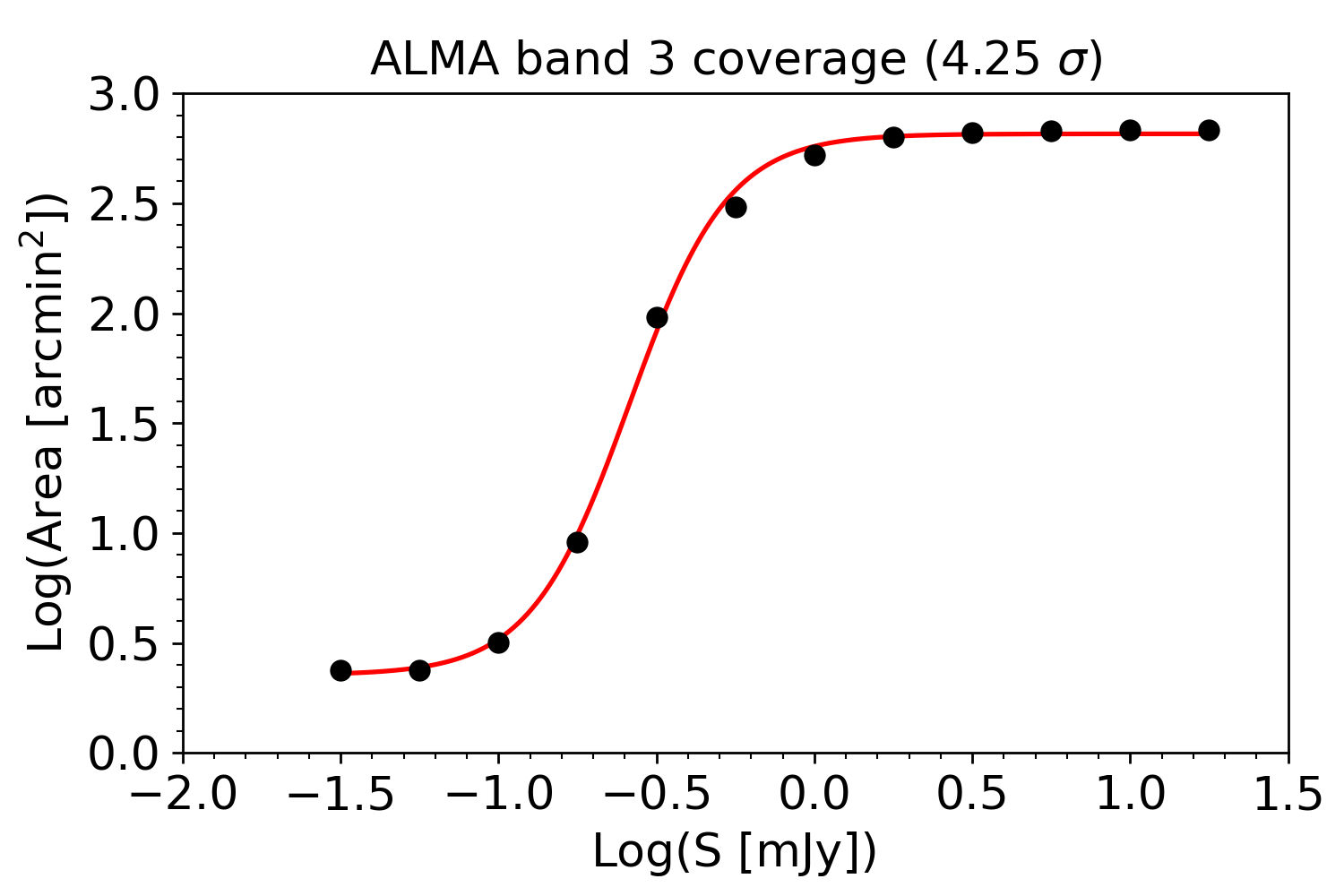}
\caption{Effective area as a function of the $4.25\,\sigma$ detection limit. The maximum value, corresponding to the brightest flux limits, amounts to $681\,\hbox{arcmin}^2$.}
 \label{fig:area}
  \end{center}
\end{figure}

\section{The sample}\label{sec:sample}

A detailed description of the calibration and imaging procedures for the ALMACAL data is provided by \citet{Oteo2016} and \citet{Chen2023}. In the following, we provide a summary of key information. The ALMACAL images were derived from ALMA’s calibration data, following a standard calibration pipeline using CASA. Two cycles of self-calibration were applied, with phase-only solutions in the first cycle and both amplitude and phase corrections in the second. Imaging was performed using natural weighting and the \citet{Hogbom1974} algorithm. Automated masking (auto-multithresh) was employed during cleaning (it was stopped when the peak emission dropped below the 2$\sigma$ noise level of the residual image, see \citealt{Chen2023}). Source fitting was carried out in the image plane.

We used an up-to-date version of the ALMACAL sample (see \citealt{Chen2023}). Since we are interested in extragalactic  sources, we used only band 3 fields at  Galactic latitude $|b| \ge 10^\circ$, thus excluding the part of the sky most heavily populated by Galactic sources. {Following the criteria detailed in \cite{Baronchelli2024}, we excluded images affected by asymmetrical and/or peculiar flux distributions, therefore selecting calibrator images characterized by a Gaussian distribution of the underlying noise. Additionally, only images exhibiting spatially uniform noise across the field (within a 10\% tolerance) were included. Finally, we excluded images characterized by a FWHM (of the synthesized beam) larger than 5''. Only 606, out of a total of 824 fields, satisfied all these criteria. In these fields, the typical value for the FWHM is $\sim$1.3'' and for the rms (at the center of the images, i.e. without the primary beam correction) is $\sim$64$\mu$Jy.}

In our analyses we considered, for each calibrator, the deepest image available\footnote{As explained e.g. in \citealt{Bonato2018}, observations of calibrators are made for every ALMA science project, several times during the same execution block (EB) or once per EB, depending on what kind of calibration is used. In this study, for each calibrator, we used the observation with the largest integration time (and in accordance with the criteria previously explained)} after the selection process. 
Following \citet{Oteo2016} and \citet{Klitsch2020}, we limited the search for sources to radii $\le 1.5\,\hbox{FWHM}_{\rm primary beam}/2$. This limiting radius corresponds to 21\% of the peak sensitivity.
On the other hand, we masked the central part of each map up to a distance of 4 arcsec (or 1 FWHM of the synthesized beam when larger than 4 arcsec) from the calibrator, to minimize the impact of possible contamination due to calibrator extensions (e.g., emissions from AGN jets and lobes).



{After applying our image selection and after masking the innermost and outermost parts of each calibrator image,} the total unmasked area {of the survey} amounts to $681\,\hbox{arcmin}^2$.
The {effective} area decreases with decreasing detection limit both because fields have different depths and because the rms noise increases with increasing distance from the center of each field. Figure\,\ref{fig:area} shows the {overall} effective area\footnote{Note that, at each flux, the effective area used to normalize our counts is computed considering circular coronas sorrounding the calibrators, each of which is characterized by its own specific sensitivity.} as a function of the flux detection limit. {For this computation, we considered as detected only sources characterized by signal to noise ratio (SNR) $>4.25\,\sigma$}\footnote{Following the methodology described in \citet{Baronchelli2024}, we set a detection threshold of 4.25\,$\sigma$ to balance sensitivity and purity. It provides a good compromise, maximizing the number of detected sources while maintaining a manageable level of contamination. The resulting contamination fraction is estimated to be around 50\%. Furthermore, this threshold roughly coincides with the minimum SNR at which contaminants can be visually distinguished.}. The fitting function writes:
\begin{equation}\label{eq:eff_area}
\log(\rm Area/\rm arcmin^{2})=\frac{a}{1+e^{-b [ \log(\rm S/ \rm mJy)-c ] }}+d,
\end{equation}
with $\hbox{a}\simeq 2.462$, $\hbox{b}\simeq 6.391$, $\hbox{c}\simeq-0.586$, and $\hbox{d}\simeq 0.354$.

The area is large enough to allow us to extend by a factor of $\simeq 30$, down to sub-mJy levels, the counts of radio sources. 

\section{Methodology }\label{sec:methodology}

\begin{table}
\caption{Band 3 detections at $\geq 4.25\,\sigma$.  $S_{\rm band 3}$ is the total flux density as given by SExtractor's FLUX\_AUTO. $\sigma_{\rm S_{\rm band 3}}$ is the associated uncertainty, including a calibration error of 5$\%$ of the total flux density (see \citealt{Bonato2018}). Total flux density and the associated uncertainty are primary beam corrected. We calculated the component of the uncertainty related to the noise (not the calibration error) with the widely used SExtractor software, which assumes that each pixel is independent of the others. However, any image reconstructed from interferometric data using a well-sampled synthesized beam (e.g., N pixels per beam) will exhibit spatially correlated noise. While in our case noise correlation represents a minor contribution, ESSENCE (\citealt{Tsukui2023}) represents an alternative method capable of handling correlated noise for any given aperture, making it a suitable reference. The SNR measures are computed on apertures as large as the beam; the root-mean-square (rms) noise is computed on apertures randomly distributed across the image. The SNR quoted does not simply correspond to the ratio between flux density and associated uncertainty as the SNR is measured in circular beam size apertures, while the flux is measured in automatic beam-shaped apertures (generally larger than the beam itself).
The last column contains the weights to be assigned to each detection to compute the number counts (see text). The $\ast$ in the last column denotes detections corresponding to sources detected by \citet{Chen2023}.}\label{tab:Catalog}
\resizebox{\columnwidth}{!}{\begin{tabular}{rrrrl}
\hline
\multicolumn{1}{c}{RA$\pm \sigma_{\rm RA}$} & \multicolumn{1}{c}{DEC$\pm \sigma_{\rm DEC}$} & \multicolumn{1}{c}{S$_{\rm band 3}\pm \sigma_{\rm S_{\rm band 3}}$} & \multicolumn{1}{c}{SNR} & \multicolumn{1}{c}{weight}  \\
\multicolumn{1}{c}{${\rm [deg]}$} & \multicolumn{1}{c}{${\rm [deg]}$} &  \multicolumn{1}{c}{${\rm [mJy]}$}  & & \\
\hline
10.051290 $\pm$ 0.000052  &  1.427770 $\pm$ 0.000052  &  1.294 $\pm$ 0.097  &  7.5  &  0.15  \\
14.238190 $\pm$ 0.000049  &  16.414420 $\pm$ 0.000049  &  1.079 $\pm$ 0.070  &  4.4  &  0.04  \\
14.451250 $\pm$ 0.000014  &  30.354350 $\pm$ 0.000014  &  17.912 $\pm$ 0.913  &  8.4  &  0.03  \\
18.076110 $\pm$ 0.000010  &  -66.579010 $\pm$ 0.000010  &  2.005 $\pm$ 0.126  &  10.5  &  0.03$\ast$  \\
19.366800 $\pm$ 0.000078  &  14.304310 $\pm$ 0.000078  &  1.123 $\pm$ 0.071  &  4.3  &  0.14  \\
27.455880 $\pm$ 0.000031  &  18.946980 $\pm$ 0.000031  &  1.163 $\pm$ 0.080  &  4.5  &  0.03  \\
34.452940 $\pm$ 0.000040  &  1.749040 $\pm$ 0.000040  &  0.641 $\pm$ 0.058  &  5.3  &  0.02  \\
34.454910 $\pm$ 0.000045  &  1.745430 $\pm$ 0.000045  &  0.776 $\pm$ 0.066  &  4.7  &  0.01  \\
37.219970 $\pm$ 0.000019  &  -3.625520 $\pm$ 0.000019  &  0.735 $\pm$ 0.045  &  4.4  &  0.01  \\
37.802360 $\pm$ 0.000011  &  -47.769180 $\pm$ 0.000011  &  0.852 $\pm$ 0.063  &  4.6  &  0.05  \\
45.860460 $\pm$ 0.000018  &  -24.118660 $\pm$ 0.000018  &  1.903 $\pm$ 0.165  &  5.0  &  0.02  \\
48.172810 $\pm$ 0.000122  &  1.553560 $\pm$ 0.000122  &  1.468 $\pm$ 0.109  &  4.9  &  0.05  \\
54.126710 $\pm$ 0.000007  &  32.306450 $\pm$ 0.000007  &  3.148 $\pm$ 0.168  &  11.5  &  0.08$\ast$  \\
54.867150 $\pm$ 0.000103  &  -1.777190 $\pm$ 0.000103  &  1.586 $\pm$ 0.092  &  4.8  &  0.10  \\
60.771900 $\pm$ 0.000014  &  25.999970 $\pm$ 0.000014  &  0.379 $\pm$ 0.032  &  4.3  &  0.02  \\
62.909290 $\pm$ 0.000031  &  -51.827210 $\pm$ 0.000031  &  0.705 $\pm$ 0.048  &  9.5  &  0.02$\ast$  \\
72.972320 $\pm$ 0.000027  &  -46.885530 $\pm$ 0.000027  &  0.273 $\pm$ 0.019  &  4.8  &  0.40  \\
74.260890 $\pm$ 0.000019  &  -23.413810 $\pm$ 0.000019  &  1.017 $\pm$ 0.088  &  4.7  &  0.02  \\
77.507530 $\pm$ 0.000032  &  18.006140 $\pm$ 0.000032  &  0.737 $\pm$ 0.075  &  4.8  &  0.03  \\
79.944660 $\pm$ 0.000145  &  -45.776760 $\pm$ 0.000145  &  2.217 $\pm$ 0.128  &  4.5  &  0.02  \\
83.163890 $\pm$ 0.000034  &  7.546800 $\pm$ 0.000034  &  0.700 $\pm$ 0.063  &  4.8  &  0.01  \\
84.706770 $\pm$ 0.000007  &  -44.087250 $\pm$ 0.000007  &  4.581 $\pm$ 0.240  &  14.3  &  0.03  \\
84.708660 $\pm$ 0.000011  &  -44.085110 $\pm$ 0.000011  &  2.833 $\pm$ 0.157  &  8.8  &  0.03  \\
85.189580 $\pm$ 0.000095  &  -54.301850 $\pm$ 0.000095  &  0.888 $\pm$ 0.071  &  7.7  &  0.02  \\
116.397560 $\pm$ 0.000036  &  10.193450 $\pm$ 0.000036  &  2.203 $\pm$ 0.128  &  4.4  &  0.02  \\
123.858280 $\pm$ 0.000034  &  36.591280 $\pm$ 0.000034  &  1.552 $\pm$ 0.089  &  13.3  &  0.02  \\
145.068650 $\pm$ 0.000114  &  26.050070 $\pm$ 0.000114  &  1.199 $\pm$ 0.074  &  5.4  &  0.03  \\
153.696730 $\pm$ 0.000005  &  23.022440 $\pm$ 0.000005  &  1.354 $\pm$ 0.080  &  5.0  &  0.22  \\
156.188220 $\pm$ 0.000022  &  19.207790 $\pm$ 0.000022  &  0.341 $\pm$ 0.036  &  4.6  &  0.15  \\
161.234940 $\pm$ 0.000034  &  6.932300 $\pm$ 0.000034  &  0.427 $\pm$ 0.046  &  4.4  &  0.04  \\
162.021670 $\pm$ 0.000082  &  -19.166960 $\pm$ 0.000082  &  8.140 $\pm$ 0.475  &  4.4  &  0.03  \\
162.035450 $\pm$ 0.000078  &  -19.156070 $\pm$ 0.000078  &  3.430 $\pm$ 0.202  &  4.6  &  0.02  \\
165.531710 $\pm$ 0.000034  &  -44.065380 $\pm$ 0.000034  &  1.720 $\pm$ 0.100  &  4.9  &  0.02  \\
166.923890 $\pm$ 0.000024  &  -30.720360 $\pm$ 0.000024  &  2.189 $\pm$ 0.124  &  4.8  &  0.01  \\
166.927020 $\pm$ 0.000026  &  -30.723500 $\pm$ 0.000026  &  1.236 $\pm$ 0.090  &  4.4  &  0.01  \\
166.931840 $\pm$ 0.000024  &  -30.730010 $\pm$ 0.000024  &  0.888 $\pm$ 0.077  &  4.9  &  0.02  \\
166.936660 $\pm$ 0.000027  &  -30.724680 $\pm$ 0.000027  &  0.730 $\pm$ 0.087  &  4.3  &  0.01  \\
166.938610 $\pm$ 0.000026  &  -30.727540 $\pm$ 0.000026  &  0.725 $\pm$ 0.072  &  4.5  &  0.02  \\
\hline
\end{tabular} }
\end{table}

\addtocounter{table}{-1}
\begin{table}
\caption{...continued.}
\resizebox{\columnwidth}{!}{\begin{tabular}{rrrrl}
\hline
\multicolumn{1}{c}{RA$\pm \sigma_{\rm RA}$} & \multicolumn{1}{c}{DEC$\pm \sigma_{\rm DEC}$} & \multicolumn{1}{c}{S$_{\rm band 3}\pm \sigma_{\rm S_{\rm band 3}}$} & \multicolumn{1}{c}{SNR} & \multicolumn{1}{c}{weight}  \\
\multicolumn{1}{c}{${\rm [deg]}$} & \multicolumn{1}{c}{${\rm [deg]}$} &  \multicolumn{1}{c}{${\rm [mJy]}$}  & & \\
\hline
166.939640 $\pm$ 0.000027  &  -30.729730 $\pm$ 0.000027  &  1.001 $\pm$ 0.083  &  4.3  &  0.02  \\
187.554780 $\pm$ 0.000019  &  25.301130 $\pm$ 0.000019  &  0.368 $\pm$ 0.035  &  4.5  &  0.02  \\
189.931750 $\pm$ 0.000051  &  -10.387180 $\pm$ 0.000051  &  1.926 $\pm$ 0.159  &  4.6  &  0.01  \\
192.113240 $\pm$ 0.000073  &  -45.987280 $\pm$ 0.000073  &  7.027 $\pm$ 0.393  &  5.8  &  0.01  \\
192.122270 $\pm$ 0.000082  &  -46.004800 $\pm$ 0.000082  &  5.606 $\pm$ 0.325  &  5.1  &  0.02  \\
192.125080 $\pm$ 0.000078  &  -45.992130 $\pm$ 0.000078  &  2.596 $\pm$ 0.244  &  5.4  &  0.04  \\
196.182110 $\pm$ 0.000011  &  -3.766190 $\pm$ 0.000011  &  1.440 $\pm$ 0.090  &  4.5  &  0.07  \\
196.182140 $\pm$ 0.000010  &  -3.769350 $\pm$ 0.000010  &  0.555 $\pm$ 0.044  &  5.2  &  0.21  \\
196.183870 $\pm$ 0.000011  &  -3.769370 $\pm$ 0.000011  &  0.853 $\pm$ 0.064  &  4.6  &  0.30  \\
199.798600 $\pm$ 0.000017  &  -12.292440 $\pm$ 0.000017  &  0.325 $\pm$ 0.031  &  4.3  &  0.03  \\
199.911040 $\pm$ 0.000033  &  -0.834300 $\pm$ 0.000033  &  1.654 $\pm$ 0.137  &  4.4  &  0.01  \\
199.912890 $\pm$ 0.000034  &  -0.835300 $\pm$ 0.000034  &  2.104 $\pm$ 0.148  &  4.3  &  0.01  \\
203.234160 $\pm$ 0.000089  &  2.009680 $\pm$ 0.000089  &  4.556 $\pm$ 0.253  &  4.3  &  0.06  \\
205.522280 $\pm$ 0.000018  &  -20.858020 $\pm$ 0.000018  &  0.449 $\pm$ 0.042  &  4.4  &  0.04  \\
215.074370 $\pm$ 0.000028  &  -6.696650 $\pm$ 0.000028  &  0.518 $\pm$ 0.060  &  4.3  &  0.02  \\
216.731600 $\pm$ 0.000028  &  -2.258460 $\pm$ 0.000028  &  0.429 $\pm$ 0.047  &  4.4  &  0.04  \\
218.250960 $\pm$ 0.000089  &  -18.025500 $\pm$ 0.000089  &  1.369 $\pm$ 0.077  &  4.5  &  0.03  \\
229.174870 $\pm$ 0.000150  &  0.260210 $\pm$ 0.000150  &  2.895 $\pm$ 0.157  &  4.4  &  0.01  \\
229.416760 $\pm$ 0.000121  &  -24.366450 $\pm$ 0.000121  &  2.370 $\pm$ 0.140  &  4.5  &  0.02  \\
239.166960 $\pm$ 0.000124  &  -33.045070 $\pm$ 0.000124  &  3.087 $\pm$ 0.189  &  5.2  &  0.02  \\
239.475690 $\pm$ 0.000057  &  -0.025870 $\pm$ 0.000057  &  2.353 $\pm$ 0.123  &  4.5  &  0.02  \\
241.899370 $\pm$ 0.000023  &  -33.521650 $\pm$ 0.000023  &  0.602 $\pm$ 0.048  &  4.3  &  0.02  \\
246.524200 $\pm$ 0.000012  &  -29.859550 $\pm$ 0.000012  &  5.189 $\pm$ 0.276  &  9.2  &  0.03  \\
253.039380 $\pm$ 0.000026  &  -4.014900 $\pm$ 0.000026  &  0.694 $\pm$ 0.083  &  4.3  &  0.02  \\
263.258060 $\pm$ 0.000051  &  -13.080310 $\pm$ 0.000051  &  2.251 $\pm$ 0.257  &  5.0  &  0.06  \\
263.258120 $\pm$ 0.000003  &  -13.080250 $\pm$ 0.000003  &  3.729 $\pm$ 0.195  &  22.7  &  0.03  \\
263.259980 $\pm$ 0.000059  &  -13.084310 $\pm$ 0.000059  &  3.199 $\pm$ 0.295  &  4.3  &  0.01  \\
263.262730 $\pm$ 0.000016  &  -13.077750 $\pm$ 0.000016  &  0.771 $\pm$ 0.065  &  4.6  &  0.18  \\
263.263460 $\pm$ 0.000014  &  -13.077480 $\pm$ 0.000014  &  0.963 $\pm$ 0.085  &  5.0  &  0.07  \\
263.264070 $\pm$ 0.000016  &  -13.078290 $\pm$ 0.000016  &  1.915 $\pm$ 0.125  &  4.6  &  0.17  \\
279.351040 $\pm$ 0.000008  &  -33.320550 $\pm$ 0.000008  &  1.444 $\pm$ 0.089  &  10.5  &  0.03  \\
304.764650 $\pm$ 0.000012  &  -52.030110 $\pm$ 0.000012  &  0.234 $\pm$ 0.024  &  4.5  &  0.31$\ast$  \\
309.155640 $\pm$ 0.000076  &  -28.511170 $\pm$ 0.000076  &  0.438 $\pm$ 0.041  &  4.7  &  0.02  \\
314.078580 $\pm$ 0.000125  &  -47.254600 $\pm$ 0.000125  &  0.898 $\pm$ 0.051  &  4.4  &  0.03  \\
314.118960 $\pm$ 0.000024  &  -58.340870 $\pm$ 0.000024  &  1.692 $\pm$ 0.128  &  4.6  &  0.02  \\
314.121090 $\pm$ 0.000025  &  -58.335770 $\pm$ 0.000025  &  3.737 $\pm$ 0.250  &  4.4  &  0.02  \\
314.122830 $\pm$ 0.000023  &  -58.339890 $\pm$ 0.000023  &  0.933 $\pm$ 0.120  &  4.9  &  0.04  \\
315.246030 $\pm$ 0.000118  &  -29.557590 $\pm$ 0.000118  &  1.481 $\pm$ 0.088  &  4.8  &  0.04  \\
316.249240 $\pm$ 0.000026  &  -48.818160 $\pm$ 0.000026  &  1.813 $\pm$ 0.135  &  4.3  &  0.02  \\
317.390870 $\pm$ 0.000042  &  -41.174290 $\pm$ 0.000042  &  0.433 $\pm$ 0.040  &  5.2  &  0.08  \\
328.672030 $\pm$ 0.000073  &  17.466630 $\pm$ 0.000073  &  0.610 $\pm$ 0.057  &  4.7  &  0.01  \\
331.916230 $\pm$ 0.000067  &  -53.783460 $\pm$ 0.000067  &  2.281 $\pm$ 0.124  &  4.5  &  0.22$\ast$  \\
331.932770 $\pm$ 0.000056  &  -53.788150 $\pm$ 0.000056  &  4.255 $\pm$ 0.226  &  5.3  &  0.02  \\
331.948360 $\pm$ 0.000067  &  -53.770400 $\pm$ 0.000067  &  2.725 $\pm$ 0.164  &  4.5  &  0.02  \\
334.718170 $\pm$ 0.000004  &  -3.595350 $\pm$ 0.000004  &  11.681 $\pm$ 0.597  &  20.9  &  0.04  \\
336.026520 $\pm$ 0.000061  &  -11.446800 $\pm$ 0.000061  &  0.702 $\pm$ 0.042  &  4.3  &  0.03  \\
338.152830 $\pm$ 0.000056  &  11.728330 $\pm$ 0.000056  &  0.686 $\pm$ 0.070  &  4.6  &  0.02  \\
338.159150 $\pm$ 0.000056  &  11.725650 $\pm$ 0.000056  &  2.399 $\pm$ 0.144  &  4.6  &  0.15  \\
343.298280 $\pm$ 0.000035  &  32.596620 $\pm$ 0.000035  &  0.637 $\pm$ 0.046  &  4.4  &  0.27  \\
344.527160 $\pm$ 0.000022  &  -27.975460 $\pm$ 0.000022  &  1.138 $\pm$ 0.118  &  4.3  &  0.01  \\
353.560640 $\pm$ 0.000029  &  7.606730 $\pm$ 0.000029  &  1.803 $\pm$ 0.170  &  4.3  &  0.01  \\
\hline
\end{tabular}}
\label{tab:Catalog}
\end{table}

\begin{table*}
\caption{Associations of our ($\geq 4.25\,\sigma$) band 3 detections with those in other ALMA bands in the ALMACAL sample. $S$ are the total, not the peak, flux densities. The uncertainties include a calibration error of 5$\%$ of the total flux density (see \citealt{Bonato2018}). Sep$_{3-\rm x}$ is the angular distance between band-3 and band-x positions. We didn't find any counterparts in the ALMA bands 8, 9, and 10.}
\centering
\resizebox{\textwidth}{!}{\begin{tabular}{cccccccccccccccc}
\hline
RA & DEC & S$_{\rm band 3}$ & $\sigma_{\rm S_{\rm band 3}}$  & S$_{\rm band 4}$ & $\sigma_{\rm S_{\rm band 4}}$ & Sep$_{\rm 3-4}^{(1)}$  & S$_{\rm band 5}$ & $\sigma_{\rm S_{\rm band 5}}$ & Sep$_{\rm 3-5}$  & S$_{\rm band 6}$ & $\sigma_{\rm S_{\rm band 6}}$ & Sep$_{\rm 3-6}$  & S$_{\rm band 7}$ & $\sigma_{\rm S_{\rm band 7}}$ & Sep$_{\rm 3-7}$ \\
${\rm [deg]}$ & ${\rm [deg]}$ &  ${\rm [mJy]}$ & ${\rm [mJy]}$ &  ${\rm [mJy]}$ & ${\rm [mJy]}$ & ${\rm [arcsec]}$ &  ${\rm [mJy]}$ & ${\rm [mJy]}$ & ${\rm [arcsec]}$ &  ${\rm [mJy]}$ & ${\rm [mJy]}$ & ${\rm [arcsec]}$ &  ${\rm [mJy]}$ & ${\rm [mJy]}$ & ${\rm [arcsec]}$ \\
\hline
18.07611 & -66.57901 & 2.0051 & 0.126  &  /  &  /  &  /  &  /  &  /  &  /  &  /  &  /  &  /  &  0.334  &  0.029  &  0.19  \\
123.85828 & 36.59128 & 1.5519 & 0.089  &  /  &  /  &  /  &  /  &  /  &  /  &  0.618  &  0.060  &  0.34  &  /  &  /  &  /  \\
246.52420 & -29.85955 & 5.1893 & 0.276  &  1.487  &  0.127  &  0.42  &  /  &  /  &  /  &  /  &  /  &  /  &  0.548  &  0.056  &  0.23  \\
263.25812 & -13.08025 & 3.7291 & 0.195  &  3.502  &  0.204  &  0.03  &  /  &  /  &  /  &  0.758  &  0.077  &  0.04  &  2.085  &  0.163  &  0.21  \\
317.39087 & -41.17429 & 0.4328 & 0.040  &  /  &  /  &  /  &  /  &  /  &  /  &  0.283  &  0.035  &  0.36  &  /  &  /  &  /  \\
\hline
\end{tabular}}
\label{tab:cm_ALMA}
\end{table*}

\begin{table}
\caption{Associations of our ($\geq 4.25\,\sigma$) band\,3 detections with the \citet{Gordon2021} VLASS catalogue. The VLASS central frequency is 3\,GHz. $S$ are the total, not the peak, flux densities.}
\centering
\resizebox{\columnwidth}{!}{\begin{tabular}{rrrrrrr}
\hline
\multicolumn{1}{c}{RA} & \multicolumn{1}{c}{DEC} & S$_{\rm band 3}$ & $\sigma_{\rm S_{\rm band 3}}$  & S$_{\rm VLASS}$ & $\sigma_{\rm S_{\rm VLASS}}$ & \multicolumn{1}{c}{Sep} \\
\multicolumn{1}{c}{[deg]} & \multicolumn{1}{c}{[deg]} &  \multicolumn{1}{c}{[mJy]} & \multicolumn{1}{c}{[mJy]} &  [mJy] & [mJy] & [arcsec]\\
\hline
10.05129  &  1.42777  &  1.294  &  0.097  &  12.648  &  1.163  &  0.04  \\
54.12671  &  32.30645  &  3.148  &  0.168  &  35.526  &  2.630  &  0.93  \\
156.18822  &  19.20779  &  0.341  &  0.036  &  0.711  &  0.347  &  0.36  \\
\hline
\end{tabular}}
\label{tab:cm_VLASS}
\end{table}

\subsection{Source extraction and photometry}\label{sec:extraction}

In order to extract detections in our selected calibrator images, and to measure their flux densities, we used SExtractor \citep{BertinArnouts1996}. 
SExtractor has been utilized on CLEANed interferometric images since its initial use \citep{Bondi2003}. While originally designed for optical astronomy, its speed and ease of use have also contributed to its widespread adoption in radio astronomy. \cite{Huynh2012} and \cite{Hancock2012} analyzed various source extraction algorithms applied to radio images, finding only minor differences in contamination, completeness, and measured fluxes. This was especially true for maps characterized by noise distribution similar to the theoretical conditions (Gaussian distribution). Importantly, we deliberately excluded images where the noise distribution significantly deviated from a Gaussian profile, especially those strongly impacted by relevant noise patterns.

For the estimation of the background, we considered a mesh grid with size equal to 4 times the FWHM (of the synthesized beam) expressed in pixels (BACK\_SIZE parameter in SExtractor) and a median filter with 2x2 pixels size (BACK\_FILTERSIZE=2). For the filtering of the image during the detection phase, we employed a Gaussian filter with size close to twice the FWHM specific to each ALMACAL image.

For the measurement of the total flux, we relied on the SExtractor ``AUTO'' estimate, which is the sum of background-subtracted pixel values inside a Kron-like \citep{Kron1980} ellipse centered on each source. {We set KRON\_FACT$=2.5$, meaning that the semi-axes of the elliptical aperture are defined as 2.5 times the Kron radius computed along each direction. The smallest possible aperture (min\_radius parameter) is set to 3.5 pixels\footnote{While we set this parameter to 3.5 pixels, we verified that, with only one exception, the size of the automatic apertures used is always larger than the beam size. For this reason, this parameter has basically no effects on our estimates.}. The minimum number of { adjacent} pixels above the detection threshold ($\hbox{DETECT\_THRESH}=1.2\,\sigma$) triggering detection was set, again, to be similar to the synthesized beam area: $\hbox{DETECT\_MINAREA}=2\pi(\hbox{FWHM}_a/2.355)(\hbox{FWHM}_b/2.355)$, $\hbox{FWHM}_{a,b}$ being the FWHM of the point spread function (PSF) along the minor and the major axis, in units of pixels.}

Our detection method is based on two successive steps and it is meant to mitigate the effects of spurious detections. We first extract (using SExtractor), only sources having a certain number of pixels (equal to the beam area) above 1.2\,$\sigma_{pixel}$ (here $\sigma$ is computed over single pixels). After this initial step, we considered only detections above 4.25\,$\sigma_{aper}$. In this second case, $\sigma$ is computed with the method of the ``random apertures'': many beam-size apertures are randomly located along the image and their flux distribution is used to estimate $\sigma$. A source detected at 6\,$\sigma$ means that its (beam-size) aperture flux is 6 times $\sigma_{aper}$, estimated with the method of the random apertures. Given the very low threshold imposed over single pixels, we expect basically every real source detectable at 4.25\,$\sigma_{aper}$ to be also originally extracted by SExtractor. This method eliminates possible unrealistic emissions due to noise patterns at scales smaller than the beam area. In any case, the missing sources would not represent a problem in our counts, as they are taken into account by our completeness curve (Fig.2 in \citealt{Baronchelli2024}). The completeness curve is computed using simulated images, injecting sources at various $\sigma_{aper}$, and extracting them using the same two-steps methodology just described.

\begin{table}
\centering
\caption{Band 3 (100\,GHz) Euclidean normalized differential number counts, $S^{2.5}dN/dS [{\rm Jy}^{1.5}{\rm sr}^{-1}]$, of sources detected at $4.25\,\sigma$ significance level.}
\begin{tabular}{rr}
\hline
\hline
$\log({\rm S [mJy]})$ & $\log({\rm Counts} [{\rm Jy}^{1.5}{\rm sr}^{-1}])$ \\
\hline
-0.62 & $ 0.18 ^{+ 0.68 }_{- 1.09 }$ \\
-0.38 & $ -0.19 ^{+ 0.66 }_{- 1.13 }$ \\
-0.12 & $ 0.01 ^{+ 0.5 }_{- 0.68 }$ \\
0.12 & $ 0.23 ^{+ 0.52 }_{- 0.75 }$ \\
0.88 & $ 0.6 ^{+ 0.46 }_{- 0.61 }$ \\
\hline
\hline
\end{tabular}
\label{tab:counts}
\end{table}

\begin{table}
\centering
\caption{Band 3 (100\,GHz) integral number counts per square degree, N($>$S), of sources detected at a $\ge 4.25\,\sigma$ significance level.}
\begin{tabular}{rr}
\hline
\hline
$\log({\rm S [mJy]})$ & $\log({\rm N}(>{\rm S}) [{\rm deg}^{-2}])$ \\
\hline
-0.75 & $ 2.04 ^{+ 0.65 }_{- 0.97 }$ \\
-0.5 & $ 1.55 ^{+ 0.56 }_{- 0.79 }$ \\
-0.25 & $ 1.35 ^{+ 0.49 }_{- 0.67 }$ \\
0.0 & $ 1.14 ^{+ 0.49 }_{- 0.67 }$ \\
0.25 & $ 0.87 ^{+ 0.46 }_{- 0.61 }$ \\
\hline
\hline
\end{tabular}
\label{tab:integral_counts}
\end{table}

\begin{figure*}
\begin{center}
\includegraphics[width=1.01\textwidth]
{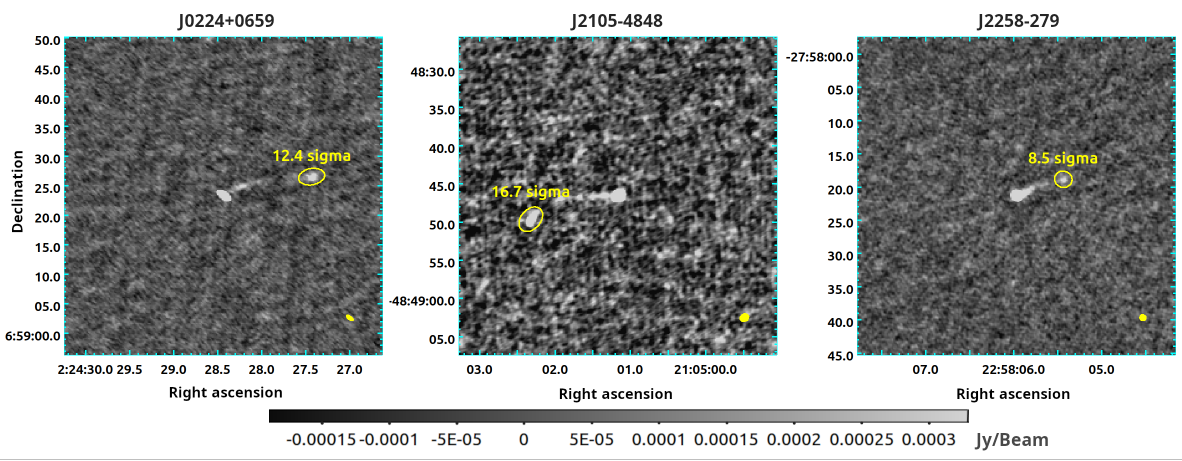}
\caption{Examples of high SNR detections {found by \emph{UMLAUT} to be contaminants for our counts (real sources but associated to the calibrators)}. In each image, the calibrator is at the center. In the bottom right of each panel, we show the size and orientation of the beam (FWHM). Left panel: the source on the right (RA=36.11429\,deg, DEC=6.99068\,deg) has a flux density of 3.09\,mJy, a $\hbox{SNR}=12.4$, and an angular distance from the calibrator of 15.4\,arcsec. Central panel: the source on the left (RA=316.25964\,deg, DEC=-48.81379\,deg) has a flux density of 5.94\,mJy, a $\hbox{SNR}=16.7$, and an angular distance from the calibrator of 11.7\,arcsec. Right panel: the source on the right (RA=344.52264\,deg, DEC=-27.97191\,deg) has a flux density of 1.55\,mJy, a $\hbox{SNR}=8.5$, and an angular distance from the calibrator of 7.7\,arcsec.}
 \label{fig:UMLAUT}
  \end{center}
\end{figure*}

\subsection{Completeness, contamination and flux boosting}

The {computation} of number counts requires delicate corrections for incompleteness and various kinds of contamination. {Here we summarize the approach we adopted.}

To estimate the completeness we resorted to simulations. After inverting {the flux in all the} images, we injected simulated sources with a 2-dimensional Gaussian PSF shape, {at} various significance levels\footnote{{This approach is based on the assumption that the average background is removed or substantially negligible across the image.}}. {By applying the same source extraction method described in Section~\ref{sec:extraction}, we could measure the} completeness as a function of the {significance (SNR) of the detections}. A simulated source was considered to be recovered {when} found within half beam size {from the position at which it was injected}.

{Using the same simulations, we could also estimate the effects of the Eddington bias \citep{Eddington1913}, also referred to as flux boosting. Positive noise fluctuations can boost faint sources above the detection threshold, making them appear brighter than they truly are. Conversely, negative fluctuations can make brighter sources appear fainter, moving them below the same threshold. Since faint sources are more numerous than bright ones, this inflates the number of detected sources.  }

{Thanks to our simulations, we can accurately quantify the fraction of sources misplaced in each flux bin. However, in order to correct the counts for the flux boosting effect, it is also necessary to know the real underlying flux density distribution of the sources. Unfortunately, this is exactly what we are trying to derive.} The way out is to exploit model predictions.

{To this end,} we used the combination of model C2Ex by \citet{Tucci2011} for radio sources with the physical model by \citet{Cai2013} for dusty star-forming galaxies (DSFGs). The C2Ex model  accurately predicted the South Pole Telescope (SPT) radio source counts \citep{Mocanu2013, Everett2020}, including those at 95\,GHz. The \citet{Cai2013} was the only one to accurately predict the SPT counts of strongly lensed DSFGs and is consistent with the 3\,mm counts by \citet{Zavala2018}, \citet{GL2019}, and \citet{Chen2023}.  {It is worth noting that while the use of theoretical models effectively allows for the correction of the flux boosting effect, the models themselves can introduce a slight secondary bias in the final number counts. This influence, however, is a lesser effect and primarily sensitive to the shape (slope) of the theoretical counts curve, rather than to the total number of sources (normalization).}

{Our choice of requiring a minimum number of adjacent pixels above the threshold for a source to be {included in the sample} reduces the number of spurious detections.  {However, this method is less effective when the noise is not random, but is spatial correlated}.

We {computed} the fraction of spurious detections as a function of the ratio between flux density {of the detected sources} and rms { (i.e., noise level)} of the images. {For this purpose, we used the same flux-inverted maps already} employed to assess the incompleteness (but without injecting simulated sources). The fraction of spurious detections corresponds to the ratio between} the number of detections in the inverted maps (assumed to be spurious) and the number of detections in the original maps. {For the flux-inverted maps, we used the same source extraction method used for the original maps, described in Section~\ref{sec:extraction}}.

{The contamination from spurious detections, as a function of the SNR,} can be represented by:
\begin{equation}
\rm Contamination \, \rm fraction=\frac{a}{1+e^{-b(\rm SNR-c)}},
\end{equation}
with a$\simeq$1.00, b$\simeq$-2.28, c$\simeq$4.34.

{The contamination curve sharply} increases below $\hbox{SNR}\simeq 5$. {On the other hand,} due to the steepness of the counts at the relevant flux densities, a decrease of the detection limit results in a more than linear increase of the number of real detections. Therefore, we push the detection limit {down} to $4.25\,\sigma$, where the contamination by spurious sources {remains lower than} $50\%$.

{Masking} the area within a 4 arcsec radius {from} the calibrator removes most of the signals {related with} the calibrator itself. {Unfortunately, some} lobes at larger angular distances {are missed by this simplistic approach. In addition,} radio sources preferentially reside in galaxy clusters and may therefore have companions whose presence biases the derivation of source counts. 

Finally, calibrators are {precisely} chosen {for being} bright sources. Consequently, the high dynamic range (peak flux / rms) of the images considered, is prone to the emergence of noise patterns and false positives at low flux densities, which are often generated during the imaging process (particularly during the ``cleaning'' phase). For the specific case of ALMA, this well known problem is implicitly described in the ALMA Science Portal's help documentation\footnote{\url{https://help.almascience.org/kb/articles/what-is-meant-by-imaging-dynamic-range}}: calibration errors and deconvolution artifacts leave structured residuals in the cleaned image, particularly around bright sources. These residuals increase the local rms, making it non-uniform across the image. Since low-flux sources are detected at a few times the local rms, if the rms is variable and underestimated, some noise peaks could be mistaken for real sources. In addition, sources in the sidelobes of the antenna beam can introduce residuals across the entire image. These structured residuals can create localized regions where random noise aligns with residual structure, mimicking faint sources. 
General analyses of the problems related with the imaging at high dynamic range can be found in \citet{Perley1999}, \citet{Uson2012}, \citet{Braun2013}. 
{The contaminants just described are due to the (strong) emission of the central calibrator and cannot be corrected using the classical contamination curve computed as previously explained.} In order to mitigate the problem, we rejected the most heavily corrupted images \citep[see e.g. Figure 1, left panel, in][]{Baronchelli2024}. However, the contaminants are still visible in many of the remaining images that we used in our analyses.

{Visually disentangling potential background sources from detections associated with the calibrator is inherently subjective and lacks reproducibility, especially at low SNR. To overcome this problem, we adopted a machine learning approach based on the K-nearest neighbor (K-NN) algorithm \citep[e.g.,][]{Altman1992}, implemented in the \emph{UMLAUT} software \citep{Baronchelli2021}. Specifically, we considered all detections with a signal-to-noise ratio (SNR) above 4.5$\sigma$ and visually classified them as either potential background sources or contaminants associated with the calibrator. Some morphological patterns make it possible to visually identify contaminants above this threshold. For example, the presence of a “bridge” connecting a calibrator to a nearby detection (see e.g. Figure~\ref{fig:UMLAUT}) or a symmetric arrangement of sources on opposite sides of a calibrator strongly suggests a physical association with the calibrator. Similarly, structured noise patterns can lead to easily identifiable spurious detections in specific regions of the image \citep[see e.g. Figures 1 and 7 in][]{Baronchelli2024}. This visually labeled dataset serves as the training set for UMLAUT, which learns the relevant classification patterns from some input parameters and applies them systematically to new detections, reducing the subjectivity inherent in manual classification.

  As an initial step, \emph{UMLAUT} determines the position of the data point under analysis in an N-dimensional ranked parameter space generated by N selected parameters. Specifically, we considered, as input parameters, those that we considered useful for our own visual inspection:
\begin{itemize}
\item{the SNR of each detection;}
\item{the aperture flux of the sources;}
\item{the distance of the detections from the calibrators (in units of FWHM);}
\item{the FWHM of the images (in arc-seconds);}
\item{the ratio between the RMS of the images computed over randomly placed beam-sized apertures and the RMS computed over single pixels\footnote{The RMS noise in an image can be computed at different spatial scales: at the pixel level (\(\sigma_{\text{pix}}\)) or over larger apertures (\(\sigma_{\text{aper}}\)). If the noise is Gaussian and uncorrelated, considering the total flux, they follow the relation \(\sigma_{\text{aper}} = \sigma_{\text{pix}} \sqrt{N}\), where \(N\) is the number of pixels in the aperture. In real observations, noise correlations may cause deviations from this simple scaling.};}
\item{the apparent size of the sources, (i.e., the ratio between the semi-major axis of the sources and the FWHM of the image's PSF);}
\item{the aperture-to-total flux ratio of the sources;}
\item{the ellipticity of the sources (i.e, the ratio between the semi-major and semi-minor axes of the sources).}
\end{itemize}	
For each detection, \emph{UMLAUT} automatically weights the input parameters using a gradient descent method that minimizes the uncertainty associated with the output.

By analyzing the closest neighbours, \emph{UMLAUT} calculates the probability of a detection to be either:
\begin{itemize}
\item{A) an actual background source or a ``normal'' spurious detection, generated by the random distribution of noise and already accounted for by the classical contamination curve, or}
\item{B) a contaminant, i.e., a foreground source or a false positive, both associated with the presence of the calibrator in the image.}
\end{itemize}
In our visual classification of the training set, we labelled the first of the two classes with ``0'' and the second class with ``1''. Consequently, the average (or distance-weighted average) of the labels associated with the closest data points, ranging from 0 to 1, can be directly used as probabilities of the detections to belong to the class of the unwanted sources (B). Therefore, the inverse of these probabilities can be used in our counts as weights for each detection, substantially removing the initial subjectivity of the visual classification, by incorporating a data-driven and objective similarity measure. }

Figure\,\ref{fig:UMLAUT} shows three examples of sources, detected {at} high SNR and located at distances substantially larger than 4\,arcsec from the calibrator. {For all of them, \emph{UMLAUT}} assigned a probability $< 0.01$ of not being associated with the central calibrator. {In these specific cases, the detections clearly correspond to real foreground sources, not to be included in our counts.}


Whether we analyze probabilities for each detection or combine them within flux bins can significantly affect the reliability of UMLAUT's results. The probability assigned to each detection is essentially an average over the closest data points in the N-dimensional parameter space. This explains why, when contaminants outnumber genuine background sources, the probability for any single detection remains low. To simplify, let's assume UMLAUT was considering the 10 closest data points in the N-dimensional parameter space (in reality this parameter can vary between 2 and 8). In this case, if a particular flux bin contains 10 detections but only one expected background source, the probability assigned to each detection will be around 0.1. Statistically, these detections are likely to be close to each other in the parameter space, meaning that none of them will individually be classified as a true background source, as their probability remains below a meaningful threshold (e.g., 0.5). Yet, we know that one of them must be real.
The way we analyze these probabilities (either individually or aggregated within flux bins) can significantly impact the reliability of UMLAUT’s results. If we evaluate each detection separately, none of them appear highly reliable, increasing the risk of misclassifications. However, the likelihood that at least one of these detections corresponds to a true background source is actually quite high when considered collectively. By grouping detections within predefined flux bins, we properly account for this cumulative probability: the combined probability (Pi$\times$Nd=0.1$\times$10=1) suggests that a genuine background source is indeed present in this group.

{To ensure reliable probability estimates from UMLAUT, we restricted its application to sources exceeding a flux density of 0.3 mJy. This decision arose from limitations in interpreting probabilities for single detections within bins containing only one (or a few) source(s). Our visual inspection revealed that two sources below this threshold were likely extensions of the calibrators, aligning with UMLAUT's low probability assignment for them being true background sources. However, such low individual probabilities within sparsely populated bins would lead to unreliable number count estimations.  Therefore, for detections fainter than 0.3 mJy, we simply relied on our visual inspection from our counts).}

\begin{figure*}
\begin{center}
\includegraphics[trim=3.0cm 0.0cm 3.0cm 0.0cm,width=0.5\textwidth]{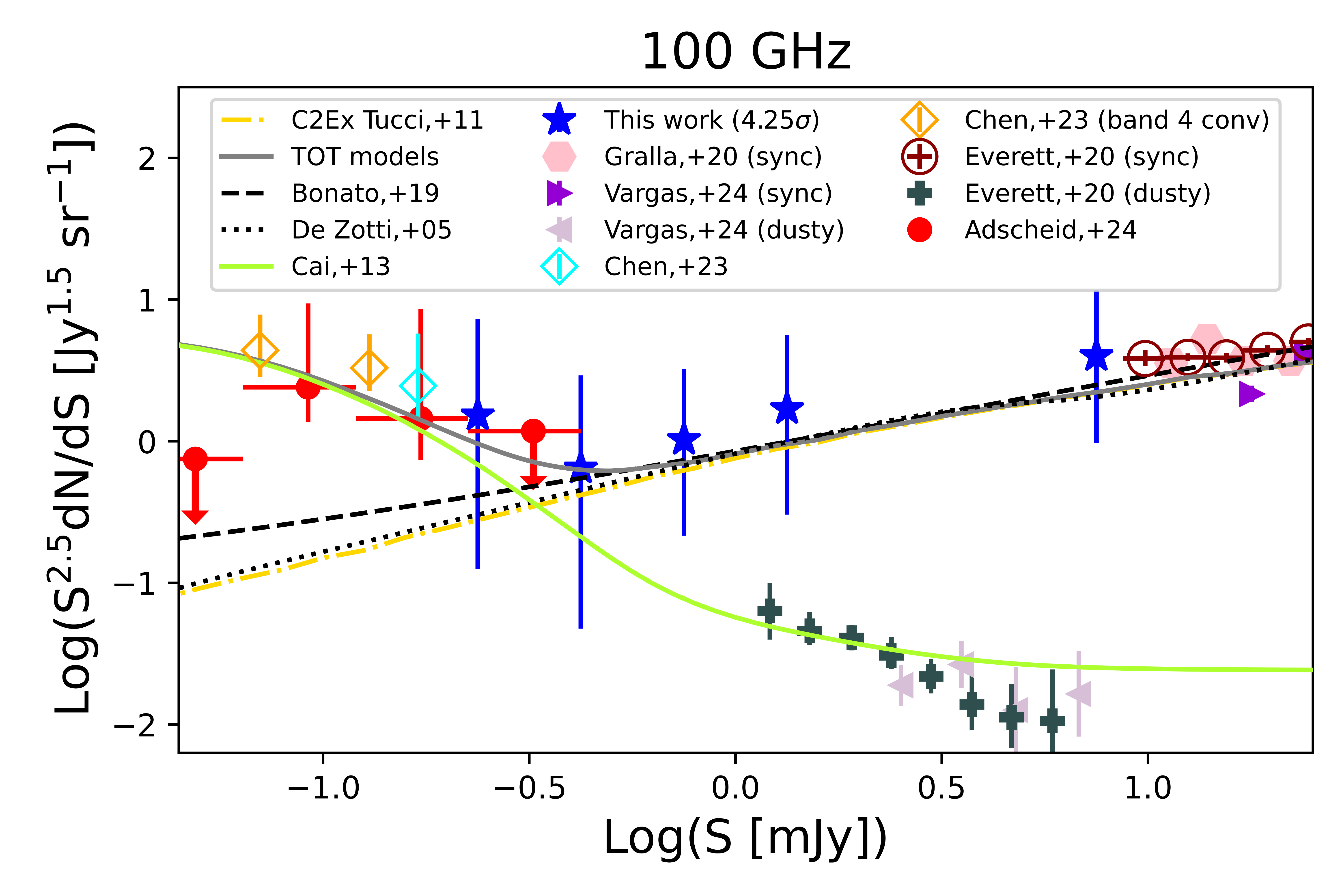}
\caption{Observational estimates of Euclidean normalized differential number counts compared with theoretical predictions. Error bars for our counts include Poisson errors and uncertainties on contamination, completeness and flux boosting corrections. Our best estimates ($4.25\,\sigma$ detection limit) are compared with the \citet{Chen2023} and the \citet{Adscheid2024} observational number counts in band\,3. We also show the band\,4 counts by \citet{Chen2023}, extrapolated to 100\,GHz using a spectral index of 3.4 ($S_\nu \propto \nu^{3.4}$), corresponding to the peak in the distribution of spectral indices between 150 and 220\,GHz of dusty galaxies reported by \citet{Everett2020}. The same spectral index was used to extrapolate to 100\,GHz the 150\,GHz counts of dusty galaxies by \citet{Vargas2024} and by \citet{Everett2020}. For the latter, we have adopted the ``no-cuts'' values extracted from Fig.\,13 of \citet{Vargas2024}. Above $\simeq 0.3\,$mJy, the counts are dominated by radio sources, in excellent agreement with predictions by \citet{DeZotti2005}, \citet[][C2Ex model]{Tucci2011}, and \citet{Bonato2019}. Our counts smoothly connect to the SPT counts by \citet{Everett2020} and the Atacama Cosmology Telescope (ACT) counts by \citet{Gralla2020} and \citet{Vargas2024}. No correction was applied to the 95\,GHz SPT counts, while the 150\,GHz ACT counts were extrapolated to 100\,GHz using the mean spectral index between 95 and 150\,GHz (-0.6) found by \citet{Everett2020} for synchrotron sources.}
 \label{fig:END_nc}
  \end{center}
\end{figure*}

\begin{figure*}
\begin{center}
\includegraphics[trim=3.0cm 0.0cm 3.0cm 0.0cm,width=0.5\textwidth]{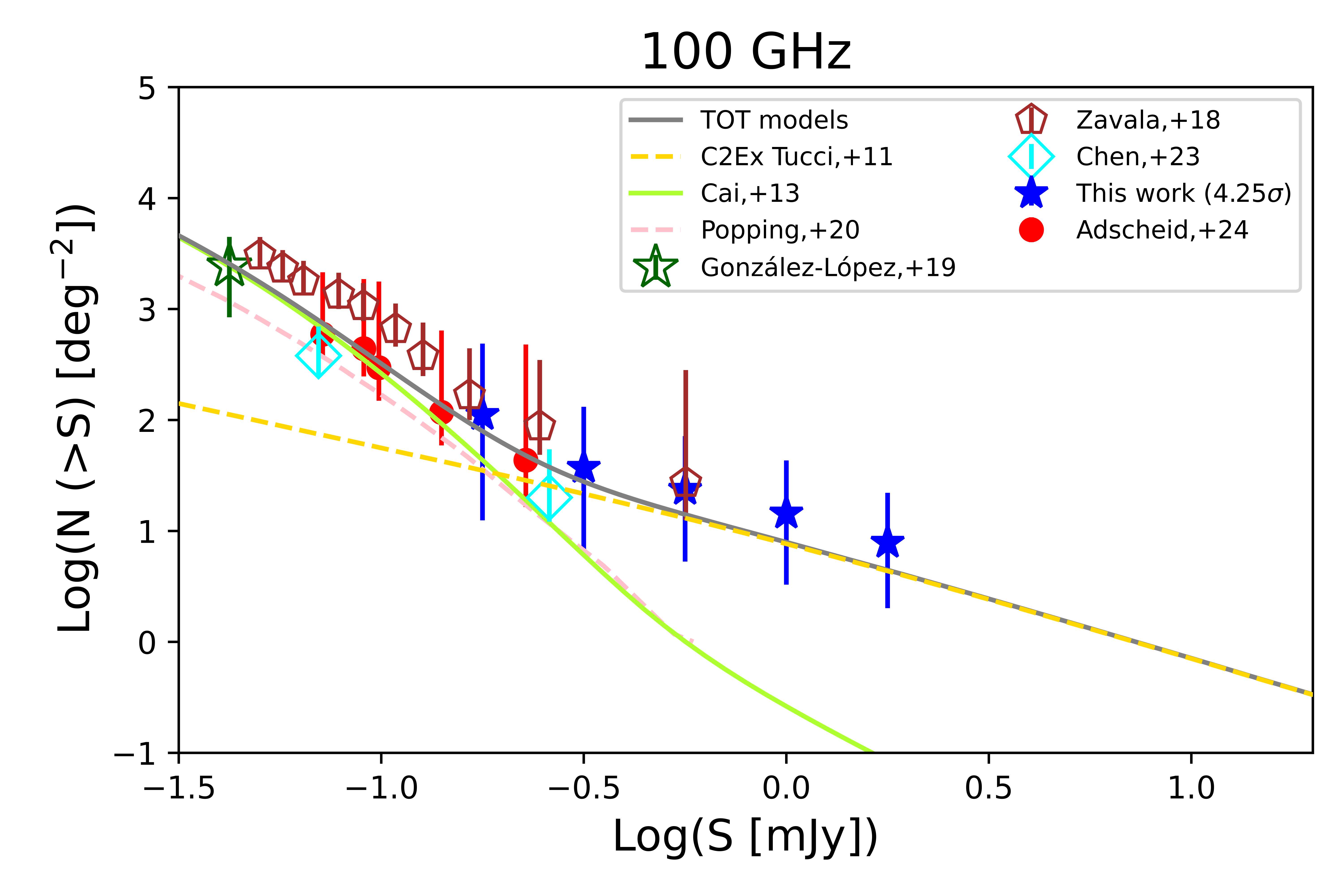}
\caption{Our integral number counts of sources detected at $\ge 4.25\,\sigma$ compared with observational estimates by \citet{Zavala2018}, \citet{GL2019}, \citet{Chen2023}, \citet{Long2024}, and \citet{Adscheid2024}; the latter were extracted from their Fig.\,8. The 2\,mm counts by \citet{Long2024} were extrapolated to 100\,GHz adopting a spectral index of 3.4 (see caption to Fig.\,\ref{fig:END_nc}). The uncertainties on our counts include Poisson errors and uncertainties on contamination, completeness and flux boosting corrections.  Also shown are predictions by \citet[][model C2Ex]{Tucci2011} for radio sources and by \citet{Cai2013} and \citet{Popping2020} for DSFGs; the latter counts were taken from Fig.\,10 of \citet{Chen2023}. The solid grey line is the sum of radio counts with those by \citet{Cai2013}.}
 \label{fig:I_nc}
  \end{center}
\end{figure*}

\section{Results}\label{sec:results}

Table\,\ref{tab:Catalog} lists the coordinates, with their errors, of detections at $\geq 4.25\,\sigma$, their flux densities given by SExtractor's FLUX\_AUTO estimate, the associated errors {(including a reference calibration error of 5$\%$; see \citealt{Bonato2018}), and the SNR. The last column contains the {overall weight factor associated with each detection}, which includes the corrections for completeness, contamination {due to} noise fluctuations, flux boosting, and the probability, yielded by \emph{UMLAUT}, that {the detection considered is not a  foreground source or in any way associated with the central calibrator. It's important to note that this weight must not be misinterpreted as the probability of a source being real.  We list only the 89 detections with $\hbox{weight}>0.01$, the contribution to the counts of those with lower weights being negligible.}

The area, $A$, over which a peak with a flux density $\ge S$ can be detected at $\ge 4.25\,\sigma$ is obtained from eq.\,(\ref{eq:eff_area}). The integral counts, $N(\ge S)$ ($\hbox{deg}^{-2}$), are the sum of $\hbox{weight}/A$ for all sources with flux density $\ge S$. The differential counts are the sum of such values within a flux density bin, divided by the bin width, $\Delta S$.

The corrections that we applied to our counts are listed here below:
\begin{itemize}
\item{A) Contamination from stochastic (i.e., Gaussian) noise distribution. This ``classical contamination'' is measured extracting sources from simulated void maps (we used the inverted maps).}
\item{B) Completeness. This curve is measured utilizing the same simulations used to estimate the contamination. In this case, we injected simulated sources and, after the extraction, we measured the fraction of undetected sources at various flux levels.}
\item{C) Flux boosting. For every simulated source, we measured the difference between injected and recovered flux.}
\item{D) Contamination from false positives and from real foreground sources. In order to measure this type of contamination we employed UMLAUT.}
\end{itemize}
We initially combined the raw completeness curve and the flux boosting effect into a single curve. This can be done using the ``C2Ex'' model by \citet{Tucci2011}. We used this theoretical model to determine the fraction of real sources that shift from one flux bin to another due to the flux boosting phenomenon, modifying the original completeness curve. This computation was performed for each individual source, as each source has its own flux and is detected at a specific SNR. The method is detailed in \citet{Baronchelli2024} (Section 2.5). By combining these two effects, we obtained a completeness curve corrected for flux boosting.
At this point, the weight associated with each detection (at each specific flux and rms value) was given by: $W1 = (1 - Contamination) / Completeness$. This weight, however, did not account for false positives (not due to the stochastic distribution of noise) or real foreground sources physically associated with the central calibrator. To correct for this effect, we used UMLAUT, as previously described. UMLAUT outputed a weight W2, which represents the probability that a source is a potential background source (i.e., the opposite of being a false positive or a real foreground source physically associated with the central calibrator).
Therefore, the final weight of each source was given by: $W3 = W1 \times W2$.

The total effective number of sources, sum of weights of detections in Table\,\ref{tab:Catalog}, is $\simeq 6$. The number of real sources is however substantially larger. To check the source reality and get hints on their nature (radio sources or DSFGs) we looked for counterparts in other ALMA bands of the ALMACAL sample using the same source extraction procedure used for band\,3 (see Sect.\,\ref{sec:extraction}).  The adopted search radius was the sum in quadrature of three times the positional errors computed as $\sigma_{{\rm pos}}=\hbox{FWHM}_{{\rm beam}}/(2\,\hbox{SNR})$, following \citet{Hughes1998}.  We found counterparts for 5 out of the 89 band\,3 detections. They are listed in Table\,\ref{tab:cm_ALMA}, where we {report, for each of them,} the total (FLUX\_AUTO) flux densities with their uncertainties and the angular distances between positions. 

We also found 3 counterparts in the Very Large Array Sky Survey (VLASS) Quick Look catalogue at 3\,GHz \citep[][Table\,\ref{tab:cm_VLASS}]{Gordon2021}. Six out of the eight sources with counterparts in other wavebands have $\hbox{weight}<0.1$; the remaining two have $\hbox{weight}\simeq 0.15$. All of them have frequency spectra indicative of being radio sources, consistent with the fact that they have $S_{\rm band3}> 0.3\,$mJy, i.e. in the range where radio sources dominate the number counts.

We stress that our approach is strictly statistical and aimed at obtaining a robust determination of the number counts. We do not consider individual detections in Table\,\ref{tab:Catalog} as true serendipitous sources; only a small fraction of them are expected to be such.

Our estimate of the Euclidean normalized differential number counts of sources detected at $\ge 4.25\,\sigma$ is displayed in Fig.\,\ref{fig:END_nc} and listed in Table\,\ref{tab:counts}. The error bars include Poisson errors and uncertainties on contamination, completeness and flux boosting corrections, obtained through simulations. 

Our differential counts bridge the flux density gap of about 1.5 orders of magnitude between the DSFG-dominated counts by \citet{Chen2023} and \citet{Adscheid2024}, and the radio-AGN-dominated counts by \citet{Everett2020}, \citet{Gralla2020} and \citet{Vargas2024}. 
The integral counts are shown in Fig.\,\ref{fig:I_nc} and reported in Table\,\ref{tab:integral_counts}. 

We also find consistency with the models by \citet{DeZotti2005}, \citet[][C2Ex model]{Tucci2011} and \citet{Bonato2019} for radio sources, and of \citet{Cai2013} and \citet{Popping2020} for DSFGs.

\subsection{Comparison with \citet{Chen2023}}

The multi-band ALMA survey by \citet{Chen2023}, exploiting ALMACAL fields, includes band 3. Their online band-3 catalogue contains 64 detections with peak $\hbox{SNR}>5$. Eight were confirmed DSFGs (type 1, according to their classification code); one of them has a peak flux density below the threshold (0.07\,mJy) chosen by them to estimate the number counts. The 44 sources classified as ``synchrotron'' (type 2) were disregarded. The 12 unclassified sources (type 0) were randomly selected as type 1 or type 2 sources based on the abundances of the confirmed type 1 and type 2 sources.

Only 34 of sources in that catalogue lie within the areas where we performed the source detection. Five of them are included in our Table\,\ref{tab:Catalog}, labeled with an $\ast$ in the last column; two are among those with the highest UMLAUT weight (0.22 and 0.31). Seven more were excluded because, at our visual inspection, looked associated to the calibrator in various ways (jets/hot spots of the calibrator, ghost images, companions). The remaining sources, including 2 out of the 7 used by \citet{Chen2023} to derive the number counts, are below our SNR threshold or have $\hbox{weight}< 0.01$. The simulations carried out both by \citet{Chen2023} and by us take care of the corrections for incompleteness and spurious detections. 

The FLUX\_AUTO estimates reported in our Table\,\ref{tab:Catalog} are systematically higher, by roughly a factor of 2, than the ``Aperture flux''  by \citet{Chen2023}. This is due to the fact that the ``Aperture flux'' always assumes the theoretical PSF of an unresolved point source, while we find that a substantial fraction of the flux comes from outside the synthesized beam size. {This must not necessarily be attributed to sources being extended. In fact, other effects arising during the imaging process (such as bandwidth smearing and time-average smearing) may alter the expected shape of the detections, still conserving the integrated flux density \citep[see e.g.,][]{Bridle1989}}. 

We computed the flux of the five sources also detected by \citet{Chen2023} using apertures of different sizes. We performed the same computation for the five calibrators in the same images. In Figure~\ref{fig:Growth_curves} the curves of growth (i.e., how the measured flux increases at increasing aperture size) are shown for the faint sources (red) and for the calibrators in the same images (green). Each curve is normalized to the maximum flux measured among the apertures considered. For this comparison, we normalized the size of the apertures to the nominal beam size in each image. 

It is clear that the growth curves are basically identical for all five calibrators. This behavior is expected because the calibrators are characterized by fluxes several hundred to thousands of times higher than the noise in the images. However, the growth curves of the faint sources (only a few sigma above the noise) significantly differ from those expected from the calibrators. Specifically, using a fixed aperture and a single aperture correction estimated from the calibrators' growth curve leads to underestimated fluxes.
Apertures sized at 2 beam diameter capture, on average, 40\% of the flux from the faint sources we are considering, whereas they capture approximately 80\% for the corresponding calibrators. All of the curves converge above a size of 4 beams. 

This data confirms that using fixed 2-beam diameter apertures strongly underestimates the flux of the faint sources.  In this context, even a Gaussian fit of the flux profile may underestimate the total flux, especially if a fixed width is considered (for example, with the intent of trying to mitigate the effect of the noise on the fit itself). Figure~\ref{fig:Growth_curves} clearly illustrates that faint sources are characterized by non-Gaussian deformed shapes, with distinct cutoffs in the wings of the faint sources profiles.

The distinct behavior of faint and bright sources underscores the importance of using variable apertures, instead of fixed ones, to accurately measure the total flux, even when those sources are expected to be unresolved. This is especially true when dealing with sources detected only a few sigma above noise, in images characterized by high dynamic range. 

\begin{figure}
\begin{center}
\includegraphics[width=\columnwidth]{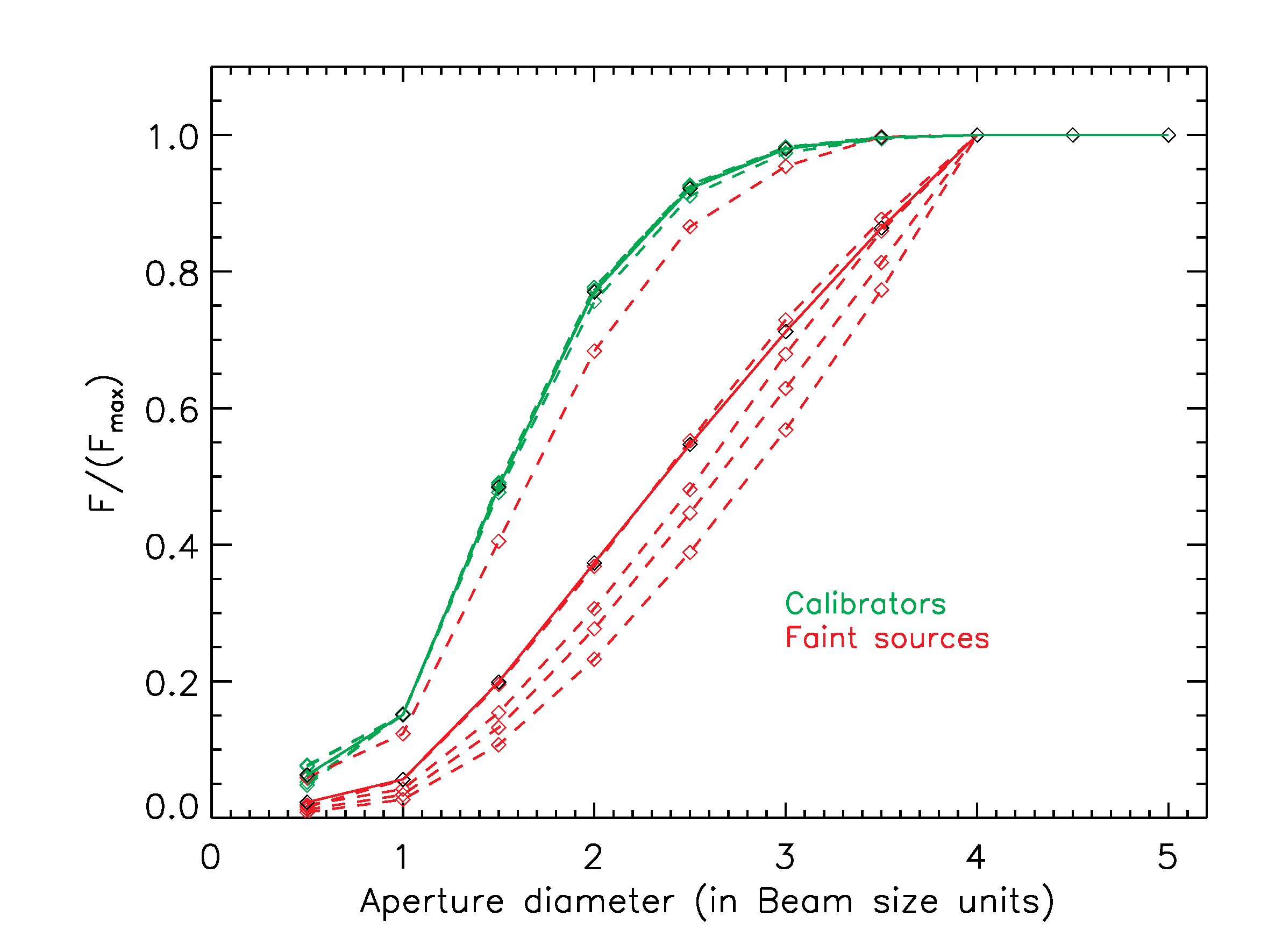}
\caption{Aperture flux as a function of the aperture size (i.e., curves of growth) for the five sources detected in both \citet{Chen2023} and this work (dashed red lines). The same function is measured also for the 5 calibrators in the same images (dashed green lines). The average curves are shown with solid lines. Each curve is normalized to the maximum flux reached among the apertures considered. The size of the apertures is normalized to the nominal beam size in each image.}
 \label{fig:Growth_curves}
  \end{center}
\end{figure}

\section{Conclusions}\label{sec:conclusions}

In addition to the complex corrections for flux boosting, incompleteness, and spurious detections that need to be applied to determine the (sub-)mm number counts, the exploitation of ALMA calibrator fields to carry out deep surveys requires dealing with biases associated {with} the presence of the calibrator itself. This is particularly relevant in the lowest frequency ALMA bands, where the radio emission is strongest. It is usual to take care of signals from the calibrator (jets, lobes) by masking the region within a few arcsec {from the calibrator itself} \citep[e.g.,][]{Klitsch2020, Chen2023}. This {approach} neglects the fact that calibrators, like radio sources in general, frequently reside in over-dense regions. {In any case, we} found that masking {the} central region {surrounding the calibrator} is insufficient: a significant {number} of high-SNR detections, related in some way to the calibrator, {are found well outside the central regions of these images. These contaminants can either be real sources or false positives generated during the imaging phase. In both  cases, such contaminants cannot be taken into account using the classical contamination correction function.}

To address this issue, we have presented the first application to the determination of number counts of a novel approach which allowed us to assign to each detection a probability of not being associated to the calibrator in any way (real physical foreground sources or false positives due to the presence of the calibrator itself in the same image). Table\,\ref{tab:Catalog} lists all detections at $\ge 4.25\,\sigma$ for which such probability is $\ge 0.01$. Further taking into account corrections for incompleteness, flux boosting and reliability of the detection we have obtained a weight measuring the contribution of each peak to the number counts.

Apart from the new methodology, the main results of this paper are the extension by a factor of about 30 (in flux density) of the 100\,GHz counts of radio sources and the best sampling compared to previous results in the region of the dominant population transition (DSFGs to radio AGNs). 
The results are in good agreement with model predictions.

\section*{Acknowledgements}
We thank Jianhang Chen and Ian Smail for their valuable comments and suggestions. MB acknowledges support from the INAF minigrant “A systematic search for ultra-bright high-z strongly lensed galaxies in Planck catalogues”. MM acknowledges support from the INAF minigrant “SHORES”

\bibliographystyle{aasjournal}
\bibliography{biblio} 

\end{document}